\documentstyle[titlepage]{article}
\newcommand{\R}{{\rm I\kern-2pt R}}

\newtheorem{proposition}{Proposition}[section]

\newtheorem{remark}{Remark}[section]
\newtheorem{definition}{Definition}[section]

\begin{document}
\begin{center}
{\bf On The Exponential of Matrices in $su(4)$\\} 
Viswanath Ramakrishna $\&$ Hong Zhou \\ 
Department of Mathematical Sciences
and Center for Signals, Systems and Communications\\
University of Texas at Dallas\\
P. O. Box 830688\\
Richardson, TX 75083 USA\\
email: vish@utdallas.edu 
\end{center}

\begin{abstract}
This note provides explicit techniques to compute the exponentials
of a variety of anti-Hermitian matrices in dimension four. 
Many of these formulae can be written down directly from the
entries of the matrix. Whenever any spectral calculations are required,
these can be done in closed form. In many instances only 
$2\times 2$ spectral calculations are required.
These formulae cover a wide variety
of applications. Conditions on the matrix which
render it to admit one of three minimal polynomials are also given.
Matrices with these minimal 
polynomials admit simple and tractable representations for their exponentials.
One of these is the Euler-Rodrigues formula.  
The key technique is the relation between real $4\times 4$ matrices
and the quaternions.

{\it PACS Numbers: 03.65.Fd, 02.10.Yn, 02.10.Hh}
\end{abstract}

\section{Introduction}
Finding the exponential of a $4\times 4$ anti-Hermitian matrix explicitly
is a problem which is of importance to quantum physics and its applications,
especially to quantum optics, quantum information processing and computation.
The problem of computing the solution to
\[
\dot{U} = iH(t)U, U(0) = I_{4}, U\in U(4)
\]
with $H(t)$ a $4\times 4$ Hermitian matrix, arises in the study of four
-level (or ``two-qubit") systems. The solution to this problem can
be reduced to the problem of computing the exponential ${\mbox exp}(i\gamma (t)
\tilde{H})$ with $\tilde{H}$ a $4\times 4$ Hermitian matrix, typically
different from $H(t)$, and $\gamma (t)$ some function. This reduction
is achieved via either a passage to a rotating frame,
approximations such as the rotating wave approximation,
or techniques such as the Wei-Norman expansion or
the Magnus expansion, or a combination thereof, 
\cite{aprraui,wei,magnusi}.
Further, in the context of controlling  four-level quantum systems,  
it is known that the unitary generators obtainable by 
allowing arbitrary time-varying external fields is precisely 
those obtainable by using only piecewise constant fields. 
This is the ``controllability with admissible controls  
is equivalent to controllability with piecewise constant controls"
result of \cite{hector}, valid for any compact Lie
group. Now determining the unitary generator after the application
of a constant field to a four-level system is a matter of 
exponentiating $4\times 4$ anti-Hermitian matrices. Further impetus
to this question is provided by the issue of universality in quantum
computation. Indeed,
in quantum computation, due to various universality type results, it
is known that to synthesize any quantum circuit it suffices to realize
unitary matrices which are the tensor products of the identity matrix
and unitary matrices which have either size two or four [see, for
instance, \cite{brylinski}]. This, in turn,
is equivalent to generating unitary matrices of size two or four
{\it selectively} (this is essentially what the identity matrices in
the tensor product amount to - the identity refers to the fact that
the external interaction which seeks to address certain levels or
qubits, does not disturb the other qubits, i.e., the addressing
of the target qubits or levels is selective). 
In particular, this requires the {\it accurate computation} of exponential
of anti-Hermitian matrices of size two or four.

The purpose of this note, keeping the above goal in mind,
is twofold. First we point out that the formulae in \cite{expistruc}
extend in a straightforward manner to provide
explicit closed-form formulae for the exponentials
of a variety of matrices
in $su(4)$ - the Lie algebra of $4\times 4$ anti-Hermitian matrices
with null trace. These formulae already cover a wide variety of
physical applications.  Secondly, we characterize when a matrix in $su(4)$
admits either a quadratic minimal polynomial or a Euler-Rodrigues
type formula for its exponential. In either instance the exponential
of the matrix has a particularly simple representation.

It is obvious that there is no loss of generality in assuming that 
the matrix being exponentiated has zero trace. It is noted
further that, in most instances,  
in the problem of exponentiating $\sum_{i=1}^{p}\gamma (t) \tilde{H},
\tilde{H} 
\in su(4)$, one can assume $\gamma$ to be constant. To illustrate
this consider the well-known formula, ${\mbox exp}(ia(t)
I_{2}\otimes \sigma_{x} + ib(t)I_{2}\otimes \sigma_{y}
+ ic(t)I_{2}\otimes \sigma_{z}) = \cos(\lambda (t))I_{4}
+\frac{\sin (\lambda (t)}{\lambda (t)}
(ia(t) I_{2}\otimes \sigma_{x} + 
ib(t)I_{2}\otimes \sigma_{y} + ic(t)I_{2}\otimes \sigma_{z}),
\lambda (t) = \sqrt{a(t)^{2} + b(t)^{2} + c(t)^{2}}$. This formula would
not suffer any modifications, beyond $\lambda (t)$ being autonomous,
if each of $a(t), b(t), c(t)$ were constant.
In particular all the results of Section 3 extend verbatim to the case
where the $\gamma (t)$ is not constant.

The formulae provided in this note rely on an associative algebra
isomorphism between $H\otimes H$ and $gl (4, R)$, with $H$ standing
for the skew-field of the quaternions and $gl (4, R)$ representing
$4\times 4$ real matrices. This isomorphism, known from the theory
of Clifford Algebras \cite{pertii}, 
has only recently been used in concrete (numerical)
linear algebra questions. To the best of our knowledge the innovative
work of \cite{haconi,ni,nii,niii} on the eigenvalue
problem for a variety of structured real matrices is the first such instance.

In \cite{expistruc} the same isomorphism was used to compute closed-form
formulae for exponentials of structured real matrices. This is indeed,
the point of departure for this note.  Given the close relationship
between the quaternions and the Pauli matrices,
it seems plausible that 
the basis for $gl(4, R)$ provided 
by the aforementioned associative algebra isomporphism is essentially
the basis for $u(4)$   
provided by the various Kronecker products
of the Pauli matrices and $I_{2}$. 
It is tempting to believe that this correspondence is as elementary as
assigning, for instance, the elements $i\otimes 1, j\otimes 1, k\otimes 1$
in $H\otimes H$ to the matrices
$\sigma_{x}\otimes I_{2}, \sigma_{y}\otimes I_{2}, \sigma_{z}\otimes 1$
etc., However, a moment's reflection shows it cannot be this simple.
In light of this, it is a pleasant circumstance that the
aforementioned plausible connection is indeed valid.
The precise correspondence is presented in the next section.

It is worth mentioning that the results presented here can also be used
to exponentiate matrices in $su(3)$. Indeed, one has to just embed such
a matrix as a principal submatrix in a $4\times 4$ matrix, with the rest
of the $4\times 4$ matrix consisting of zero entries.  
A different
application would be to exponentiate matrices in $so(6, R)$ (the Lie algebra
of $6\times 6$ real anti-symmetric matrices) matrices. $SU(4)$ is a double
cover of $SO(6, R)$, \cite{pertii}. By making 
this covering homomoprhism explicit,
one can reduce the problem of exponentiation in $so(6,R)$ to finding
exponentials in $su(4)$.   

The balance of this note is organized as follows. In the next section,
the relation between the $H\otimes H$ basis for $gl(4, R)$ and the
Pauli tensor product basis is made explicit. The same section also
establishes notation used throughout this note. The third section
presents $su(4)$ analogues of the results of \cite{expistruc}.
In particular, several illustrations drawn from 
important applications are given.  
The fourth section presents conditions which ensure that a $su(4)$ 
matrix either has a quadratic minimal polynomial or admits a Euler-
Rodrigues' formula. The same section also presents conditions equivalent
to a $su(4)$ matrix to stem from a normal matrix (i.e., if $X = B + iC$,
then $BC = CB$, where $B, C$ are the skew-symmetric and symmetric parts
of $ X\in su(4)$, respectively). The final section offers conclusions. 
\section{Notation and Preliminary Observations}
The following definitions and notations will 
be frequently met in this work: 
\begin{itemize}
\item $gl(n, R)$ represents the algebra of
real
$n\times n$ matrices. This is, of course, the Lie algebra of
the Lie groups of real invertible matrices. 
\item $SU(n)$ represents the Lie group
of $n\times n$ unitary matrices of determinant
one. $su(n)$ represents the corresponding Lie algebra
of $n\times n$ skew-Hermitian, traceless matrices.
$su(2)\otimes su(2)$ is the Lie algebra spanned by matrices
of the form $I_{2}\otimes U + V\otimes I_{2}, U, V\in su(2)$.
Note it is customary to use the terminology ``anti-Hermitian" for
skew-Hermitian matrices.
\item $R_{n}$ represents the matrix with $1$ on the anti-diagonal and
zeroes elsewhere. 
Any $n\times n$ matrix $A$, satisfying $A^{T}R_{n} + R_{n}A = 0$
said to be {\it perskewsymmetric}.     
{\it Persymmetric} matrices are those matrices, $X$, which satisfy
$X^{T}R_{n} = R_{n}X$. 
Such matrices are symmetric about the anti-diagonal. 
  
\item $J_{2n}$ is the $2n\times 2n$ matrix which, in block form,  
is given by $J_{2n} = \left ( \begin{array}{cc}
0_{n} & I_{n}\\
-I_{n} & 0_{n}
\end{array}
\right )$. 
Matrices, $Z$, satisfying
$Z^{T}J_{2n} = J_{2n}Z$ are called {\it skew-Hamiltonian} (sometimes
anti-Hamiltonian). The term ``Hamiltonian" will not typically be used
in the sense of quantum mechanics, unless specified to
the contrary (i.e., it will not be used
to mean a Hermitian matrix). 

\item Throughout $H$ will be denote the field (more precisely
the division algebra) of the {\it quaternions},
while $P$ stands for the
{\it purely imaginary} quaternions, tacitly identified with $R^{3}$.
Further, in this note the symbol $i$ will be used for both the
corresponding complex number and the corresponding quaternion.
The context should make it clear which of the two is implied.  
                                                                                
\end{itemize}

\begin{remark}
\label{minipolyi}
\begin{itemize}
\item i) {\rm Throughout this note, use of the following observation will
be made: Let $X$ be an $n\times n$ matrix satsifying $X^{2} + c^{2}I_{n}
= 0, c\neq 0$. Then $e^{X} = \cos (c) I_{n} + \frac{\sin (c)}{c}X$.
Here $c^{2}$ is allowed to be complex, and $c$ is then taken to be
$\sqrt{r} e^{i\frac{\theta}{2}}$, with $c^{2} = re^{i\theta},
\theta \in [0, 2\pi)$.} 
\item {\rm The fact that any matrix which satisfies $X^{3} = 
-c^{2}X, c\neq 0$,
satisfies $e^{X} = I + \frac{\sin (c)}{c}X + \frac{1-\cos (c)}{c}X^{2}$
(the  Euler -
Rodrigues's formula) will also be used. Once again $c^{2}$ is permitted
to be complex. Note that any matrix which satisfies $X^{2} + c^{2}I_{n}
= 0, c\neq 0$, automatically satisfies $X^{3} =
-c^{2}X, c\neq 0$. For such matrices the exponential formula in i) is
better to work with than the Rodrigues' formula. Therefore, in this note
we will allude to a } matrix admitting an Euler-Rodrigues formula only
if its minimal polynomial is of the form $x^{3} + c^{2}x$.

\item iii) {\rm Explicit formulae 
for $e^{A}$ can be produced if the minimal polynomial
of $A$ is known and it is low in degree 
(cf., \cite{zenii} where such formulae are written down from
the characteristic polynomial). 
However, since the corresponding
explicit formulae for $e^{A}$ are more complicated than the
ones corresponding to i) and ii), they will not be pursued here.}  

\end{itemize}
\end{remark}    

\vspace*{2.7mm}
                                                                                
\noindent {\bf $H\otimes H$ and $gl(4, R)$}: The algebra isomorphism
between $H\otimes H$ and $gl(4, R)$, which is central to this work is
the following:

\begin{itemize} 
\item Associate to each product
tensor $p\otimes q\in H\otimes H$,
the matrix, $M_{p\otimes q}$, of the map which sends $x\in H$
to $px\bar{q}$, identifying $R^{4}$ with $H$ via the basis $\{1,i,j,k\}$.
Thus, if $p = p_{0} + p_{1}i + p_{2}j + p_{3}k; 
q = q_{0} + q_{1}i + q_{2}j + q_{3}k$, then
\[
M_{p\otimes q} = [x\mid y\mid u \mid v] 
\]
with
$x, y, u, v$, the columns of the matrix $M_{p\otimes q}$, given
by the vectors in $R^{4}$ representing the quaternions $p\bar{q},
pi\bar{q}, pj\bar{q}, pk\bar{q}$ respectively. 
Here, $\bar{q} = q_{0} - q_{1}i - q_{2}j - q_{3}k$.

\item Extend this to the full tensor product by linearity,
This yields an algebra isomorphism
between $H\otimes H$ and $gl(4, R)$. In particular, a basis for
$gl(4, R)$ is provided by the sixteen matrices $M_{e_{x}\otimes e_{y}}$
as $e_{x}, e_{y}$ run through $1, i, j, k$.
\end{itemize}

This connection, which
is known from the theory of Clifford Algebras, has been put
to great practical use in solving eigenvalue problems for structured
matrices by Mackey et al., \cite{haconi,ni,nii,niii}.
It can also be used for finding exponentials,
$e^{A}, A\in gl(4, R)$, \cite{expistruc}.

\begin{remark}
\label{canonicformi}
Canonical Form for $X\in su(4)$ : {\rm Let $X =iH$, with
$H$ Hermitian and traceless. Then
\begin{equation}
\label{fullexpansion}
H = \sum_{i=1}^{3}\alpha_{i}I_{2}\otimes\sigma_{i}
+ \sum_{i=1}^{3}\beta_{i}\sigma_{i}\otimes I_{2} + \sum_{j=1}^{3}
\sum_{k=1}^{3}\gamma_{jk}\sigma_{j}\otimes\sigma_{k}, \alpha_{i},
\beta_{i},
\gamma_{jk}\in R
\end{equation}
It is well known that via conjugation by a local unitary
transformation (ie., conjugation via a $U\in SU(2)\otimes SU(2)$)
$H$ can be put into the form
\begin{equation}
\label{canonic}
\sum_{i=1}^{3}a_{i}I_{2}\otimes\sigma_{i}
+ \sum_{i=1}^{3}b_{i}\sigma_{i}\otimes I_{2} + \sum_{i=1}^{3}
c_{i}\sigma_{i}\otimes\sigma_{i}
\end{equation}
with $a_{i}, b_{i}, c_{i}
\in R$.  We will use this canonical form at some points
 in Section 4 (but not in Section 3).
Furthermore this local unitary transformation is
determined by finding the singular value factorization of the real
$3\times 3$ matrix $(\gamma_{jk})$.
But this amounts to finding the spectral
factorization of a real $3\times 3$ symmetric matrix - which can be
performed in closed form, \cite{symmetricsing}}.
                                                                                
\end{remark}

\vspace*{2.7mm} 

\noindent {\bf Relation to the Pauli Tensor Product Basis}
As mentioned in the introduction, the above basis for $gl(4, R)$
is closely related to the basis $\sigma_{i}\otimes \sigma_{j}, i,j
= 1, \ldots, 4$(with $\sigma_{0} = I_{2}, \sigma_{1} = \sigma_{x},
\sigma_{2}
= \sigma_{y}, \sigma_{3} = \sigma_{k}$). The precise relation
is tabulated below:

\begin{tabular}{cc}
{\bf Pauli Tensor Basis} & {\bf Quaternion Tensor Basis} \\
$I_{2}\otimes I_{2}$ &  $M_{1\otimes 1}$\\
$\sigma_{x}\otimes I_{2}$ & $M_{i\otimes k}$\\
$\sigma_{y}\otimes I_{2}$ & $-iM_{1\otimes j}$\\ 
$\sigma_{z}\otimes I_{2}$ & $M_{i\otimes i}$\\
$I_{2}\otimes\sigma_{x}$ & $M_{k\otimes j}$\\
$I_{2}\otimes\sigma_{y}$ & $iM_{i\otimes 1}$\\ 
$I_{2}\otimes\sigma_{z}$ & $M_{j\otimes j}$\\
$\sigma_{x}\otimes\sigma_{x}$ & $M_{j\otimes i}$\\
$\sigma_{x}\otimes\sigma_{y}$ & $-iM_{1\otimes k}$\\
$\sigma_{x}\otimes \sigma_{z}$ & $-M_{k\otimes i}$\\
$\sigma_{y}\otimes\sigma_{x}$ & $-iM_{1\otimes k}$\\
$\sigma_{y}\otimes\sigma_{y}$ & $M_{i\otimes j}$\\
$\sigma_{y}\otimes\sigma_{z}$ & $iM_{j\otimes 1}$\\
$\sigma_{z}\otimes\sigma_{x}$ & $-M_{j\otimes k}$\\
$\sigma_{z}\otimes\sigma_{y}$ & $-iM_{1\otimes i}$\\
$\sigma_{z}\otimes\sigma_{z}$ & $M_{k\otimes k}$ 
\end{tabular}

\section { Some Closed Form Formulae for Exponentials in $su(4)$}
In this section, we provide closed form formulae for the
exponentials of several matrices in $su(4)$, {\it without resorting
to the canonical form in Equation (\ref{canonic})}. 
These formulae are based on expressing the
matrix in question as a sum of commuting summands, each of which satisfies
the condition in i) of Remark (\ref{minipolyi}). These formulae can be 
divided
into two classes: i) those which can be directly written down from the
entries of the matrix; ii) those that require the spectral factorization
of an associated real $3\times 3$ symmetric matrix. This latter spectral
factorization can be achieved in closed form, \cite{symmetricsing}.
In particular, for several cases only a $2\times 2$ spectral factorization
is needed. These will be pointed out. Since most of these formuale are the
$su(4)$ analogues of the results in \cite{expistruc}, proofs will be
provided only for cases not considered in \cite{expistruc}. 
In the interests of brevity, we have not considered analogues of every
possible result in \cite{expistruc}.

\begin{remark}
\label{conventionii}
{\rm Consider $X\in su(4)$, written as $X = B + iC$, with $B, C$ real.
Suppose it is skew-Hamiltonian, for instance. Then a simple calculation
reveals that the real matrices $B, C$ are skew-Hamiltonian as well.
Hence so is the real matrix $B + C$.  
This observation yields the $H\otimes H$ representation of such an
$X\in su(4)$. The basic properties used in exponentiating the corresponding
real matrix $B + C$ in \cite{expistruc} was that it could be 
expressed as the sum of commuting summands, each of which is annihilated
by a polynomial of the type in i) of Remark (\ref{minipolyi}. Now these
properties are not vitiated by the presence of the imaginary unit $i$ in $X$.
Therefore their exponentials are similarly found. The only difference
is that the hyperbolic trigonometric functions in the formula for
$e^{B + C}$ will now be replaced by their ordinary trigonometric equivalents.
Similar arguments hold if $X$ is perskewsymmetric etc.,} 
\end{remark}
   
\subsection {Exponentials Directly From the Entries}

Below a list of three families of
matrices in $su(4)$, whose exponentials can be directly found from
their $H\otimes H$ representations, is presented.  

\begin{enumerate}
\item {\bf Symmetric, Tridiagonal, $S_{ii} = 0$}
Consider \[
S = i\left ( \begin{array}{cccc}
0 & \alpha & 0 & O\\
\alpha & 0 & \beta & 0\\
0 & \beta & 0 & \gamma \\
0 & 0 & \gamma & 0
\end{array}
\right )
\]
Since such matrices arise in several applications, it is interesting
to note that they can be easily exponentiated.
Indeed, such an $S$
has the following representation
\[
S = i[M_{p\otimes i} + M_{q\otimes j} + M_{r\otimes k}]
= X + Y + Z
\]
with $p = (0, \frac{\beta}{2} , 0), q = (\frac{\beta }{2}, 
0, \frac{\gamma + \alpha }
{2}),
r = (0, \frac{\gamma - \alpha}{2} , 0), \alpha , \beta , \gamma \in R$.
In terms of the Pauli tensor basis, $S$ is
$i\frac{\beta}{2} (\sigma_{x}\otimes \sigma_{x}) + 
i\frac{\beta}{2}\sigma_{y}\otimes\sigma_{y}
+ i\frac{\gamma - \alpha}{2}I_{2}\otimes\sigma_{x} + 
i\frac{\alpha - \gamma}{2} 
\sigma_{z}\otimes\sigma_{x}$ .
Now note that $Y$ commutes with both $X$ and $Z$, while $X$ and $Z$
anticommute. Further each squares to a negative constant times the
identity. So $e^{S}$ is given
by
\[
e^{S} = [\cos (\lambda_{1})I_{4} + i\frac{\sin (\lambda_{1})}{\lambda_{1}}
(M_{p\otimes i} + M_{r\otimes k})][\cos (\lambda_{2})I_{4}
+ i\frac{\sin (\lambda_{2})}{\lambda_{2}}M_{q\otimes j}] 
\]
In terms of the Pauli matrices this becomes 
$e^{S} =  [\cos (\lambda_{1})I_{4} + i\frac{\sin (\lambda_{1})}{\lambda_{1}}
(\frac{\beta}{2} (\sigma_{x}\otimes \sigma_{x}) + \frac{\alpha - \gamma}{2} 
\sigma_{z}\otimes\sigma_{x})]
[\cos (\lambda_{2}I_{4}
+ i\frac{\sin (\lambda_{2})}{\lambda_{2}}( \frac{\beta}{2}
\sigma_{y}\otimes\sigma_{y}
+ \frac{\gamma + \alpha}{2}I_{2}\otimes\sigma_{x})]
$
with $\lambda_{1} = \frac{1}{2}\sqrt{\beta^{2} + (\gamma - \alpha )^{2}},
\lambda_{2} =  \frac{1}{2}\sqrt{\beta^{2} + (\gamma + \alpha)^{2}}$.

\item {\bf Perskewsymmetric $X$ :}
Such an $X\in su(4)$ satisfies, in addition, $X^{T}{R} = -RX$.
Such matrices are expressible in the form
\[
i[p_{1}\sigma_{z}\otimes I_{2} + p_{2}\sigma_{x}\otimes \sigma_{z} +
\alpha \sigma_{y}\otimes\sigma_{z} + q_{1}I_{2}\otimes\sigma_{z}
+ q_{2}\sigma_{z}\otimes\sigma_{x} + \beta\sigma_{z}\otimes\sigma_{y}]
\]
Their exponential is given by
\[
[\cos (\lambda_{1})I_{4} + i\frac{\sin (\lambda_{1}}{\lambda_{1}}
(p_{1}\sigma_{z}\otimes I_{2} + p_{2}\sigma_{x}\otimes \sigma_{z} +
\alpha \sigma_{y}\otimes\sigma_{z})]
[\cos (\lambda_{2})I_{4} + i\frac{\sin (\lambda_{2}}{\lambda_{2}}
(q_{1}I_{2}\otimes\sigma_{z}
+ q_{2}\sigma_{z}\otimes\sigma_{x} + \beta\sigma_{z}\otimes\sigma_{y})]
\]
with $\lambda_{1} = \sqrt{\mid\mid p\mid\mid^{2} + \alpha^{2}},
\lambda_{2} = \sqrt{\mid\mid q\mid\mid^{2} + \beta^{2}}$

\item {\bf Skew-Hamiltonian $X$:}
These matrices satisfy, in addition, $X^{T}J = JX$.
Such matrices are associated with time-reversal symmetries, \cite{greiner}.
More specifically, a Hamiltonian (in the usage of quantum mechanics),
$H$, i.e., a Hermitian $H$, is associated to time-reversal symmetry
if $H^{T}J = JH$. Clearly if $H$ satisfies this additional condition,
so does $X= iH$.
Such matrices are expressible in the form
\[
i[bI_{4} + p_{1}\sigma_{y}\otimes\sigma_{y} + p_{2}I_{2}\otimes\sigma_{z}
+ p_{3}I_{2}\otimes \sigma_{x} + c\sigma_{z}\otimes\sigma_{y}
+ d\sigma_{x}\otimes\sigma_{y}]
\]
Their exponential is given by
\[
e^{ib}[\cos (\lambda )I_{4} + i\frac{\sin (\lambda )}{\lambda} 
(p_{1}\sigma_{y}\otimes\sigma_{y} + p_{2}I_{2}\otimes\sigma_{z}
+ p_{3}I_{2}\otimes \sigma_{x} + c\sigma_{z}\otimes\sigma_{y}
+ d\sigma_{x}\otimes\sigma_{y})],
\lambda =  \sqrt{\mid\mid p\mid\mid^{2} + c^{2} + d^{2}}
\]

\end{enumerate}

\subsection {The Purely Imaginary Case}
The following algorithm for exponentiating a matrix $X\in su(4)$,
which is simultaneously symmetric (equivalently purely imaginary)
follows directly from the corresponding algorithm for exponentiating
purely real symmetric matrices in \cite{expistruc}. The only
difference is that the $\cosh (), \sinh ()$ in \cite{expistruc}
will be replaced by $\cos (), \sin ()$.
Note that such an $S$ will not have any terms in
Equation (\ref{fullexpansion})
corresponding to members of the Pauli tensor basis, which 
contain {\it precisely one} $\sigma_{y}$ term.  

\begin{itemize}
\item Represent the given symmetric $S\in su(4)$ as the
matrix as $i[M_{p\otimes i} + M_{q\otimes j} + M_{r\otimes k}],
p,q,r\in P$. 
\item Identifying the pure quaternions $p,q,r$ with vectors in
$R^{3}$, find the spectral factorization of the real $3\times 3$
symmetric matrix $X^{T}X$, where $X = [p \mid q \mid r]$. 
Thus $X^{T}X  = \sum_{1=}^{3}\sigma_{i}^{2}v_{i}v_{i}^{T}$.
\item Compute $u_{i} = Xv_{i}$ (Note $u_{i}$ are almost
the left singular vectors. The only difference is
$\mid\mid u_{i}\mid\mid = \sigma_{i}$). Then $S = i\sum_{i=1}^{3}
M_{u_{i}\otimes v_{i}}$. Hence,
\begin{equation}
\label{expisymmetric}
e^{S} = \Pi_{i=1}^{3} (\cos (\sigma_{i})I_{4} + i\frac{\sin (\sigma_{i})}
{\sigma_{i}}M_{u_{i}\otimes v_{i}})
\end{equation} 
\end{itemize} 
\begin{definition}
\label{bisymmetric}
{\bf Bisymmetric Type:}
{\rm For several important examples only a $2\times 2$ spectral factorization
is needed (which is extremely easy to write). Since the archtypical example
is provided by a matrix in $su(4)$ which is, in addition,
bisymmetric 
(i.e., simultaneously
symmetric and persymmetric), we will, to avoid circumlocution,
call all such matrices} of the bisymmetric type.  
\end{definition}

\subsection{Illustrative Examples}
We provide some important illustrations of the formulae developed 
in this section.

\noindent {\bf Illustration 1: Rabi Oscillations in Four Level Systems}
In \cite{fujii} a detailed calculation, via a calculation of 
eigenvectors and eigenvalues, 
is provided to calculate the evolution of a four level system,
being irradiated by three laser fields, under the rotating
wave approximation and under resonance. Specifically,
they consider a four level 
system with energy levels $\{ E_{k}, k=1, \ldots ,
3\}$ which satisfy $E_{1} - E_{0} > E_{2} - E_{1} > \ldots
> E_{3} - E_{2}$. This system is irradiated by three laser fields
with frequencies $\omega_{k} = E_{k} - E_{k-1}, k=1, \ldots , 3$.
After passage to a rotating frame, and under the assumptions of 
resonance and the rotating wave approximation, the unitary generator
in the rotating frame satisfies  
\begin{equation} 
i\dot{\tilde{U}} = (E_{0}I_{4} + C)\tilde{U}
\end{equation}
with
\[
C = \left ( \begin{array}{cccc}
0 & g_{1} & 0 & 0\\
g_{1} & 0 & g_{2} & 0\\
0 & g_{2} & 0 & g_{3}\\
0 & 0 & g_{3} & 0
\end{array} \right )
\]
Here the $g_{i}$ are the amplitudes of the three laser fields.  
Thus $\tilde{U}(t) = e^{-iE_{0}t}{\mbox exp} (-iCt)$.
In \cite{fujii} ${\mbox exp} (-iCt)$ is calculated by a direct calculation
of the eigenvalues and eigenvectors of the matrix $-iC$. 
Now, $-iC$ is precisely a symmetric, tridiagonal matrix with a zero diagonal
- i.e., of the type considered in item 5) of the list.
Thus, ${\mbox exp} (-iCt)$ may be found directly and is equal to
\[
e^{-iCt} = 
= [\cos (\lambda_{1})I_{4} + i\frac{\sin (\lambda_{1})}{\lambda_{1}}
(\beta (\sigma_{x}\otimes \sigma_{x}) + (\alpha - \gamma )
\sigma_{z}\otimes\sigma_{x})]
[\cos (\lambda_{2}I_{4}
+ i\frac{\sin (\lambda_{2})}{\lambda_{2}}( \beta\sigma_{y}\otimes\sigma_{y}
+ (\gamma + \alpha)I_{2}\otimes\sigma_{x})]
\]
with $ \alpha = -\frac{1}{2}g_{1}t, \beta = -\frac{1}{2}g_{2}t,
\gamma = -\frac{1}{2}g_{3}t,
\lambda_{1} = \sqrt{\beta^{2} + (\alpha - \gamma )^{2}},
\lambda_{2} =  \sqrt{\beta^{2} + (\gamma + \alpha)^{2}}$.
                                                                                
A laborious but straightforward calculation confirms that the matrix
entries provided
by the above representation of ${\mbox exp} (-iCt)$ coincide with 
those in \cite{fujii}.

\noindent {\bf Illustration 2: Josephson Junction}
In \cite{jjunctioni,aprrauii} the following system is considered;
\[
i\dot{U} = HU, U(0) = I_{4}
\]
with
\[
H = \left ( \begin{array}{cccc}
E_{00} & -\frac{1}{2}E_{J1} & -\frac{1}{2}E_{J2} & 0\\
-\frac{1}{2}E_{J1} & E_{10} & 0 & -\frac{1}{2}E_{J2}\\
-\frac{1}{2}E_{J2} & 0 & E_{10} & -\frac{1}{2}E_{J1}\\
0 & -\frac{1}{2}E_{J2} & -\frac{1}{2}E_{J1} & E_{00} 
\end{array}
\right)
\]
In \cite{jjunctioni} $E_{00}, E_{10}, E_{J1}, E_{J2}$ are taken
to be constants reflecting current technology.
Thus $U(t) = e^{-iHt}$. Now note that 
\begin{equation}
-iH = -i [\frac{1}{2}(E_{00} + E_{10})I_{4} -
-\frac{1}{2}E_{J2}\sigma_{x}\otimes I_{2} -\frac{1}{2}E_{J1}
I_{2}\otimes \sigma_{x} + \frac{1}{2}(E_{00} - E_{10})
\sigma_{z}\otimes \sigma_{z}]
\end{equation}
In terms of the Pauli tensor basis
this is $ -i [ \frac{1}{2}(E_{00} + E_{10})M_{1\otimes 1}
-\frac{1}{2}E_{J2}M_{i\otimes k} -\frac{1}{2}E_{J1}M_{k\otimes j}
+ \frac{1}{2}(E_{00} - E_{10})M_{k\otimes k}]$ 
Hence, $e^{-iHt} = e^{-i(E_{00} + E_{10})t}e^{-i\tilde{H}t}$,
with $-i\tilde{H} = 
-i[M_{p\otimes k} + M_{q\otimes j}]$,
with the purely imaginary quaternions of the form
$p = p_{1} i + p_{3} k, q = q_{3}k$. 
Thus, the singular value factorization
of the $2\times 2$ matrix \[
\left (\begin{array}{cc}
p_{1} & 0\\ 
p_{3} & q_{3}
\end{array}
\right )
\]
has to be found. Thus, this is an example of the {\it bisymmetric
type}. Specifically, the calculations proceed as follows:
$\tilde{H} = -i[M_{u_{1}\otimes v_{1}} + M_{u_{2}\otimes v_{2}}]$     
with 
$v_{1} = \cos\theta i -\sin\theta k, v_{2} = \sin\theta i + \cos\theta k$,
Here  $\tan (2\theta ) = \frac{2p^{T}q}{q^{T}q - p^{T}p}$.
Further $u_{1} = p_{1}\cos\theta i - (p_{3} + q_{3})\sin \theta k,
u_{2} = p_{1}\sin\theta i + (p_{3} +q_{3})\cos\theta k$. 
Then $\mid\mid v_{i} \mid\mid = 1, i=1,2$,
while $\mid\mid u_{1} \mid\mid = \sigma_{1}
= \sqrt{q^{T}q\cos^{2}\theta + r^{T}r\sin^{2}\theta -
q^{T}r\sin (2\theta )}$, $\mid\mid u_{2} \mid\mid = \sigma_{2}  
=\sqrt{q^{T}q\sin^{2}\theta + r^{T}r\cos^{2}\theta +
q^{T}r\sin (2\theta )}$.
Hence,
\begin{equation}
e^{-i\tilde{H}t}
= [\cos \sigma_{1}I -i\frac{\sin\sigma_{1}}{\sigma_{1}}
M_{u_{1}\otimes v_{1}}]
[\cos \sigma_{2}I -i\frac{\sin\sigma_{2}}{\sigma_{2}}
M_{u_{2}\otimes v_{2}}]
\end{equation}
This reads, in terms of the Pauli matrices, as the following

\noindent {\bf Illustration 3. Scalar Coupling Hamiltonian}
The matrix being exponentiated is
$X = -iH = i[aI_{4} + b\sigma_{z}\otimes I_{2}
+cI_{2}\otimes \sigma_{z} + d\sigma_{z}\otimes \sigma_{z}
+ e\sigma_{x}\otimes\sigma_{x} + f\sigma_{y}\otimes\sigma_{y}]$.
This is the so-called scalar coupling Hamiltonian, and is widely
used in NMR spectroscopy. The corresponding $H\otimes H$ representation
is given by
\[
X = -i[aM_{1\otimes 1} + bM_{i\otimes i} + cM_{j\otimes j}
+ dM_{k\otimes k} + eM_{j\otimes i} + fM_{i\otimes j}]
= -i[aM_{1\otimes 1} + M_{p\otimes i} + M_{q\otimes j} 
+ M_{r\otimes k}]
\]
with $p = (b, e, 0), q= (f, c, 0),
     r = (0,0,d)$.
	     
Hence $e^{tX} = e^{iat}{\mbox exp} \  -it(M_{p\otimes i} + M_{q\otimes j}
+ M_{r\otimes k})$. Thus, it remains to find the exponential of
the purely imaginary symmetric matrix
$-it(M_{p\otimes i} + M_{q\otimes j}
+ M_{r\otimes k})$. Now notice that the real matrix $-t[p\mid q\mid r]$,
is a block diagonal matrix, with the northwest block a real
$2\times 2$ matrix and the southeast block, the $1\times 1$ matrix
$(-td)$. Hence, one needs to only find the singular value
factorization of the real
$2\times 2$ matrix
\[
-t\left (\begin{array}{cc}
b & f\\
e & c
\end{array} \right ) = [\tilde{p} \mid \tilde{q}]
\]
Hence this is also of the {\it bisymmetric type}.
The corresponding right singular vectors of
$-t[p\mid q\mid r]$, written as quaternions, are 
$v_{1} = \cos\theta i -\sin\theta j, v_{2} = \sin\theta i + \cos\theta j,
v_{3} = k$,
Here  $\tan (2\theta ) = \frac{2\tilde{p}^{T}\tilde{q}}
{\tilde{q}^{T}\tilde{q} - \tilde{p}^{T}\tilde{p}}
= \frac{2(bf +ec)}{f^{2} + c^{2} - b^{2} - e^{2}}$.
Further $u_{1} = -t(b\cos\theta i - f\sin \theta j),
u_{2} = -t(e\sin\theta i + c\cos\theta j), u_{3}
=-tdk$. Then $\mid\mid v_{i} \mid\mid = 1, i=1,2,3$,
while $\mid\mid u_{1} \mid\mid = \sigma_{1}
= t\sqrt{\tilde{p}^{T}\tilde{p}\cos^{2}\theta + 
\tilde{q}^{T}\tilde{q}\sin^{2}\theta -
\tilde{p}^{T}\tilde{q}\sin (2\theta )}$,
$\mid\mid u_{2} \mid\mid = \sigma_{2}
= t\sqrt{\tilde{p}^{T}\tilde{p}\sin^{2}\theta + 
\tilde{q}^{T}\tilde{q}\cos^{2}\theta +
\tilde{p}^{T}\tilde{q}\sin (2\theta )}, \mid\mid u_{3}\mid\mid = td$
Hence,
\[
e^{-itX} =  e^{-iat}[\cos(\sigma_{1})I_{4} 
+\frac{\sin(\sigma_{1}}{\sigma_{1}}M_{u_{1}\times v_{1}}]
[\cos(\sigma_{2})I_{4} 
+\frac{\sin(\sigma_{2}}{\sigma_{2}}M_{u_{2}\times v_{2}}]
[\cos(\sigma_{3})I_{4} 
+\frac{\sin(\sigma_{3})}{\sigma_{3}}M_{u_{3}\times v_{3}}]
\]
 
\begin{remark}
{\rm There are several other practical applications which lead
to the problem of exponentiating $su(4)$ matrices of the bisymmetric
type. Examples include superconducting circuits for solid-state
quantum computation \cite{goongi}, J cross polarization experiments
\cite{bongobandhu}, Heisenberg Hamiltonians (under the assumption
that only one of the three components of the magnetic field, assumed
to be constant in time, is active
during any period of time).} 
\end{remark} 

\begin{remark}
\label{magicbasis} 
{\rm The number of matrices which can be easily exponentiated in
this fashion can be expanded by combining the above observations
together with some useful conjugations. Two classes of such
conjugations immediately spring to mind. The first is obviously
the class of local unitary transformations. Thus, for instance
the matrix $X_{1} = i(a\sigma_{z}\otimes\sigma_{z} + b\sigma_{y}\otimes I_{2}
+ cI_{2}\otimes\sigma_{y})$ is explicitly locally unitarily
equivalent to $X_{2} = 
i(a\sigma_{z}\otimes\sigma_{z} + b\sigma_{x}\otimes I_{2}
+ cI_{2}\otimes\sigma_{x})$. The former is not a symmetric matrix,
while the latter is (in fact, it is of the bisymmetric type). However,
$e^{X_{1}}$ is easily found once $e^{X_{2}}$ is. The second type of
conjugation is via the so-called magic basis matrix (see \cite{makhlini},
for instance). Explicilty, letting
\[
V = \frac{1}{\sqrt{2}}\left ( \begin{array}{cccc}
1 & 0 & 0 & i\\
0 & i & 1 & 0\\
0 & i & -1 & 0\\
1 & 0 & 0 & -i
\end{array} \right )
\]
it is known that $V (so (4, R))V^{*} = su(2)\otimes su(2)$.
It is instructive to examine its effect on some of the other matrices
considered here. Thus, for instance
\begin{itemize}
\item $X$, symmetric, tridiagonal with $X_{ii} = 0$ implies
$VXV^{*} = i[\frac{\beta}{2}\sigma_{y}\otimes I_{2} + 
\frac{\beta}{2}\sigma_{y}\otimes \sigma_{x}
+ \frac{\gamma - \alpha}{2}\sigma_{y}\otimes\sigma_{z} + 
\frac{\alpha - \gamma}{2}\sigma_{z}\otimes\sigma_{y}]$
If one writes such a matrix explicitly, it is not
clear that it too can be written as the sum of commuting summands,
each of which are easily exponentiated. 

\item $X$, skew-Hamiltonian implies
$VXV^{*} = i[aI_{2}\otimes I_{2} + b\sigma_{y}\otimes \sigma_{x}
+c\sigma_{y}\otimes \sigma_{y} + d\sigma_{y}\otimes\sigma_{z}
+ e\sigma_{x}\otimes I_{2} + f\sigma_{z}\otimes I_{2}]$

\item $X$, perskewsymmetric implies
$VXV^{*} = i[a\sigma_{x}\otimes\sigma_{x} + b\sigma_{x}\otimes\sigma_{z}
+cI_{2}\otimes\sigma_{y} + d\sigma_{y}\otimes\sigma_{y} + e\sigma_{z}\otimes
\sigma_{y} + f\sigma_{x}\otimes I_{2}]$
\end{itemize}

Thus, all such matrices are readily exponentiated.} 
\end{remark}
  
\section{Euler-Rodrigues Formulae}
In this section we will find conditions which imply that a given
$X\in su(4)$ admits one of three minimal polynomials. For such
$X$ the corresponding formulae for $e^{X}$ is very handy.
In particular, one of them is an Euler -Rodrigues type formula,
which explains the title of the section. Finally, we will provide
conditions that an $su(4)$ matrix $X = B + iC$ is of the
normal type, i.e., $BC = CB$. Note that, since $X$ is a complex matrix,
being of the normal type is not the same as being normal.
For some of the results below we will appeal to the canonical form
for $X$ in Equation (\ref{canonic}).
This section will make use of the eigenvalue structure of matrices
in $su(4)$. However, one does not need to determine the eigenvalues
themselves.

Given an annihilating polynomial $p$ for any matrix $X$, one can
use $p$ to find $e^{X}$. There are at least two manners in which
to achieve this. One finds the zeroes of $p$ (but not the eigenvectors
of $X$), and then 
proceeds to use any of a variety of methods
(e.g, interpolation) to find $e^{X}$ \cite{hhornii}. Alternatively
one can express, via $p$,
higher order powers of $X$ in terms of lower
orders and use this to establish formulae for $e^{X}$.
In general, the representations of $e^{X}$ obtained via either method
are not always easy to work with. There are however three instances
when either method produces the same representation and this is 
particularly easy to manipulate.  Specifically, these are:  
\begin{itemize}
\item {\bf 1. Quadratic Type I:}
$p(x) = x^{2} + c^{2}$. The corresponding formula for $e^{X}$
is $e^{X} = \cos (c) I_{n} + \frac{\sin (c)}{c}X$.
\item {\bf 2. Quadratic Type II:}
$p(x) = x^{2} + 2\beta x + \gamma, \beta\neq 0$.
Now $e^{X} = e^{-\beta }[(\cos\sigma + \frac{\beta\sin(\sigma )}{\sigma})I
+ \frac{\sin\sigma}{\sigma}X]$, with $\sigma = \sqrt{\beta^{2} - \gamma}$.
\item {\bf 3. Cubic Type I:}
$p(x) = x^{3} + c^{2}x$. In this case, $e^{X}
= I + \frac{\sin (c)}{c}X + \frac{1-\cos (c)}{c}X^{2}$
\end{itemize}

\begin{remark}
{\rm The results to be presented here should be seen as a complement
to those in the previous section. In Section 3, only the first of
the above minimal polynomials was used. There will be several $X\in su(4)$
which are amenable to the techniques of either section. Consider, for
instance, $X = -iJ(t)(\sigma_{x}\sigma_{x} + \sigma_{y}\sigma_{y}
+ \sigma_{z}\sigma_{z})$.  This matrix arises in the study of quantum
dots, \cite{goongi}. The corresponding $p$ is $x^{2} - 2iJx + 3J^{2}$.
However, $X$ is also the sum of three commuting terms, each of which
is annihilated by a polynomial of the first type in the above list.}
\end{remark}

We begin with an explicit expression for the characteristic polynomial
of an $X\in su(4)$ in canonical form.

\begin{proposition}
\label{characteristicpoly}
{\rm Consider an $X\in su(4)$ in canonical form as given in Equation
(\ref{canonic}). Let its characteristic polynomial be 
$x^{4} + \mu x^{2} + \nu x + \pi$.
Then \begin{itemize}
\item i) $\mu  = 2\sum_{i=1}^{3}(a_{i}^{2}
+ b_{i}^{2} + c_{i}^{2})$
\item ii) $\nu = -8i (\sum_{i=1}^{3}a_{i}b_{i}c_{i}
 - \Pi_{i=1}^{3}c_{i})$.
\item $\pi = \frac{1}{4}\{2(\sum_{i=1}^{3}(a _{i}^{2}
+ b_{i}^{2} + c_{i}^{2}))^{2}
- 4[(\sum_{i=1}^{3}(a_{i}^{2} + b_{i}^{2} + c_{i}^{2})^{2} \\ 
+ 4 \sum_{i=1}^{3}\sum_{j=1}^{3}a_{i}^{2}b_{j}^{2} 
+ 4 \sum_{i=1}^{3}(a_{i}^{2}c_{i}^{2} + b_{i}^{2}c_{i}^{2})\\
+ 2\sum_{i,j=1; i\neq j}^{3}c_{i}^{2}c_{j}^{2}
- 4\sum_{i,j,k=1; i\neq j\neq k}^{3}a_{i}b_{i}c_{j}c_{k}]\}$ 
\end{itemize} }
\end{proposition}  

 {\it Proof:} These formulae follow from Newton's identities,
which imply that the coefficients of
the characteristic polynomial can be expressed in terms of the trace
of suitable powers of $X$, in conjunction with ${\mbox Tr}\ (X) = 0$.
Further, ${\mbox Tr}\ (X^{i}), i=2,\ldots , 4$ were calculated by
using the $H\otimes H$ representation of $X$ and looking for the
$1\otimes 1$ term in $X^{i}$. It is worth emphasizing that the ease
of quaternion multiplication renders it unnnecessary to calculate $X^{3}$
or $X^{4}$ fully. Indeed, besides calculating the $1\otimes 1$ term
in $X^{3}$, one needs to find only those terms in $X^{3}$ which
would yield a $1\otimes 1$ term in $X^{4}$ (and quaternion multiplication
facilitates this process). 

We can now give a simple characterization of when $X$'s minimal polynomial
is of either {\bf quadratic type I} or {\bf cubic type I}.

\begin{proposition}
\label{typeIs}
{\rm $X \in su(4)$ has i) minimal polynomial
$p(x) = x^{2} + c^{2}$ iff $\nu = 0$ and $\mu^{2} = 4\pi$;
ii) minimal polynomial $p(x) = x^{3} + c^{2}x$ iff $\nu = 0 = \pi$.
Furthermore, in  case i) $c^{2} = \frac{\mu}{2}$, while  in
case ii) $c^{2} = \mu$.}
\end{proposition}

{\it Proof:} First, in view of $X$'s diagonalizability,
$p(x) = x^{2} + c^{2}$ is the minimal polynomial
iff the characteristic polynomial has two distinct roots (which add up to
zero) each repeated twice. Similarly, $p(x) = x^{3} + c^{2}x$
is the minimal polynomial iff 
the characteristic polynomial has two simple distinct 
roots (which add up to zero) and a double root equal to zero.

Suppose first that $\nu = 0$. Then the characteristic polynomial 
is a quadratic for $x^{2}$. The first case occurs precisely when 
this quadratic has a double root, i.e., when $\mu^{2} = 4\pi$.
Similarly, the second case occurs when one of the roots of
this quadratic is nil, i.e., precisely when $\pi = 0$ in addition.

Conversely, suppose the minimal polynomial is $p(x) = x^{2} + c^{2}$.
Now using the characterization of the coefficients of the characteristic
polynomial in terms of the elementary symmetric functions of the eigenvalues,
it follows that $\nu = 0$ and $\mu^{2} = 4\pi$. Similarly, if
$p(x) = x^{3} + c^{2}x$ the same characterization yields $\nu = 0 = \pi$.

\begin{remark}

{\rm

\noindent i) Using these conditions it is easy to
write down examples of $X\in su(4)$ which admit genuine
Euler-Rodrigues formulae, i.e., $X$ which have cubic Type I
minimal polyynomials. For instance, $X = i(I_{2}\otimes \sigma_{x}
+ \sigma_{x}\otimes I_{2} + \sigma_{y}\otimes \sigma_{y}
+ c\sigma_{z}\otimes\sigma_{z})$, where $c$ is any real solution
of the quartic $c^{4} + 14c^{2}
- 8c + 17 = 0$. This quartic admits at least two real solutions.
Indeed, if all solutions were complex, then they must be of the form
$a +ib, a -ib, -a + id, -a -id$, since there is no $c^{3}$
term. It is easy to see that, if this is the case, then the
coefficient of $c^{2}$ has to be necessarily negative.  
Note further, that $c$, in this example, could easily
be allowed to be time-varying.

\noindent ii)
It is noted in passing that one can write down the exponential of generic
$X$ which satisfy $\nu = 0$ (i.e., those cases for which
neither of $\mu^{2} = 4\pi$ nor $\pi = 0$ hold), 
since in this case the all eigenvalues
of $X$ are distinct and the corresponding interpolation based formula
\cite{hhornii} assumes a simple form.}
\end{remark}

Characterizing when $X$ has a minimal polynomial of
{\bf Quadratic Type II} via coefficients of the characteristic
polynomials does not seem fruitful. Therefore, we provide a different
characterization. For this characterization we do not
require that $X$ be placed in the form of Equation (\ref{canonic}), though
obviously the stated conditions would simplify for $X$ in canonical form.

\begin{proposition}
\label{quadminiII}
{\rm Let $X\in su(4)$ be expressed as $M_{p\otimes 1} + M_{1\otimes q}
+ i[M_{r\otimes i} + M_{s\otimes j} + M_{t\otimes k}]$,
with $p, \ldots , t$ purely imaginary quaternions.
Denote by $C = [r\mid s\mid t]$.
Then $X$ admits $x^{2} + 2\beta x + \gamma$, with $\beta\neq 0$, as its minimum
polynomial iff there is a $\tilde{\beta}\in R$ satisfying
the following conditions:
\begin{eqnarray}
\label{likepurify}
C^{T}p &=& \tilde{\beta}q\\ \nonumber
Cq &=& \tilde{\beta}p\\ \nonumber
pq^{T} - {\mbox Co} \ (C) &=& \tilde{\beta}C
\end{eqnarray} 
where ${\mbox Co} \ (C)$ is the matrix of cofactors of $C$.
If these conditions hold, then i)
$\gamma = - [\mid\mid p\mid\mid^{2}
+ \mid\mid q\mid\mid^{2} + \mid\mid r \mid\mid^{2} + \mid\mid s\mid\mid^{2}
+ \mid\mid t\mid\mid^{2}]$; ii) $\beta = i\tilde{\beta}$.}
\end{proposition} 

{\it Proof:} From a variety of viewpoints it should be clear that
if $x^{2} + 2\beta x + \gamma$ is to be the minimum polynomial of $X$,
then necessarily $\beta$ is purely imaginary, while $\gamma$ 
is real. Using this fact, the above conditions stem 
from a direct calculation of $X^{2}$.

\begin{remark}
{\rm

\noindent The purpose of this remark is to identify some situations,
under which, the system of equations in Equation (\ref{likepurify})
admits solutions.   

\vspace*{1.8mm}

\noindent i) When $C$ has rank one, Equation (\ref{likepurify}) always
possesses solutions, i.e., it is always possible to find $p, q, \tilde{\beta}$
satisfying them for the given $C$. Such a $C$ always possesses a representation
of the form $C = uv^{T}$, with $u^{T}u = v^{T}v$ (it is easy to
find such a representation). Picking $\tilde{\beta} = u^{T}u
= v^{T}v, p =u\sqrt{\tilde{\beta}}, 
q = v\sqrt{\tilde{\beta}}$ we find that Equation (\ref{likepurify})
is satisfied. 
This yields a systematic procedure to construct examples admitting a quadratic
minimal polynomial of {\bf type II}.

\noindent ii)  Conversely, starting with a non-zero $p$
one can find $q, C$ such that Equation (\ref{likepurify})
always holds. The key to this is to beserve that
if $C$ is invertible in Proposition (\ref{quadminiII}), 
then the conditions given in Equation (\ref{likepurify}) can be written
in a different form, to wit: $CC^{T}p = \tilde{\beta}^{2}p,
q = C^{-1} (\tilde{\beta}p); \tilde{\beta}(pp^{T} - CC^{T}) = 
{\mbox det} (C) I$. This yields a method to construct more examples
of $X$ admitting a quadratic minimal polynomial of Type II. Pick a $p\neq 0$. 
Choose $\tilde{\beta} = \sqrt{1 + p^{T}p}$, and pick a $C$
satisfying ${\mbox det}(C) = -\beta$ and $CC^{T} = I + pp^{T}$. 
Finally, set $q = C^{-1} (\tilde{\beta}p)$.

This can always be achieved by picking
$C$ to be a solution of the equation $CC^{T} = I + pp^{T}$  with a negative
determinant, since $I + pp^{T}$ is positive definite and thus 
possesses a square root.
For instance, one could multiply the easily determined
unique positive definite square
root of $I + pp^{T}$ by ${\mbox diag} (1,1,-1)$ to obtain a $C$
with determinant $-\tilde{\beta}$.
Further, ${\mbox det}(CC^{T}) = 1 + p^{T}p$ and obviously
$CC^{T}p = \tilde{\beta}^{2}p$.
Indeed, the eigenvectors of $I + pp^{T}$ are $p$ (with eigenvalue
$1 + p^{T}p = \tilde{\beta}^{2}$), and any two 
vectors orthogonal to $p$ (corresponding
to eigenvalue 1 with double multiplicity).

\noindent iii) If precisely one of $p$ or $q$ is zero, then there is no
solution to Equation (\ref{likepurify}). 
When both are zero, there is a solution iff $CC^{T}$ is proportional
to the identity matrix, i.e., iff the vectors $r, s$ and $t$ are
orthogonal and have the same length. Note, in this case 
$X = i[M_{1} + M_{2} + M_{3}]$, with 
the $M_{i}$ commuting, and each with a quadratic minimal polynomial
of type {\bf I}. Further, this is precisely the case wherein the
canonical form $X$, as in Equation (\ref{canonic}), is
$X = i[c_{1}\sigma_{x}\otimes\sigma_{x} + c_{2}\sigma_{y}\otimes\sigma_{y}
+ c_{3}\sigma_{z}\otimes\sigma_{z}$, with either $c_{1} = c_{2} = c_{3}$
(in the event ${\mbox det} (C) > 0$) or $c_{1} = c_{2} = -c_{3}$     
( in the event ${\mbox det} (C) <  0$). However, one does not
require passage to this canonical form for finding $e^{X}$.

\noindent iv) Similar conditions can be written down one $p,q,r,s,t$
for $X$ to admit other minimal polynomials. We omit them in the
interests of brevity.}
\end{remark}

\noindent {\bf Conditions for ``Normality"}
Next, given $X = B + iC$, we characterize, when $[B, C] = 0$,
i.e., when the real matrix $B + C$ is normal. The motivation should
be obvious -  it is possible to exponentiate both $B$ and $iC$ in closed
form, and hence $X$. 
While the statement of this result uses the canonical form
given by Equation (\ref{canonic}), much of the proof does not
require it.

\begin{proposition}
{\rm Let $X = B + iC = M_{p\otimes 1 + 1\otimes q} + iM_{r\otimes i +
s\otimes j + t\otimes k}, p, \ldots , t\in P$ be in canonical
form. Suppose, without loss of generality, that at least one 
of $p, q$ is non-zero. 
Then $[B, C] = 0$ iff the following conditions hold:
\begin{itemize}
\item i) $p\neq 0, q = 0$: $a_{1} = a_{3} = c_{1} = c_{3} =0$
\item ii) $p\neq 0, q\neq 0$: $a_{1} = a_{3}= b_{1} = b_{3} = 0$,
$\mid \frac{b_{2}}{a_{2}}\mid = \mid \frac{c_{1}}{c_{3}}\mid = 1$
\item iii) $p =0, q\neq 0$: $b_{1} = b_{3} = c_{1} = c_{3} =0$
\end{itemize} } 
\end{proposition}

\noindent {\it Proof:} For any $X\in su(4)$ (even those not
in canonical form) a quick calculation reveals
that $[B, C] = 2[p\times r\otimes i +
p\times s\otimes j + p\times t\otimes k +
r\otimes (q \times i) + s\otimes (q\times j) + t\otimes (q\times k)]$.
If $X$ is in the canonical form in Equation (\ref{canonic}) then
we have
\[
p= ( -a_{2}, 0, 0); q = (0, b_{2}, 0); r = (b_{3}, c_{1}, 0);
s = (c_{2}, a_{3}, a_{1}), t = (b_{1}, 0, c_{3})
\]

Hence $[B, C]$ is twice the matrix representation of
$b_{1}b_{2}i\otimes i + (c_{3}b_{2} - c_{1}a_{2}) k\otimes i +
a_{1}a_{2}j\otimes j - a_{2}a_{3} k\otimes j - b_{3}b_{2}i\otimes k
+ (a_{2}c_{3} - c_{1}b_{2})j\otimes k$.
The conclusion follows from this.

\begin{remark}
{\rm By applying the proof of the previous result to $X$ not in
canonical form, one can deduce other commutativity results.
For instance, suppose $Y$ is in canonical form, and one defines
$Y_{1} = i(\sum_{i=1}^{3}a_{i}I_{2}\otimes \sigma_{i}
+\sum_{i=1}^{3}b_{i}\sigma_{i}\otimes I_{2})$ and
$Y_{2} = i\sum_{i=1}^{3}c_{i}\sigma_{i}\otimes\sigma_{i}$.
Then $[Y_{1}, Y_{2}] = 0$ iff i) $\frac{c_{3}}{c_{2}} = 
\frac{b_{1}}{a_{1}}$; ii) $\frac{c_{3}}{c_{1}}
= \frac{b_{2}}{a_{2}}$; and iii) $\frac{c_{2}}{c_{1}}
= \frac{b_{3}}{a_{3}}$.
To see this let $X = B + iC = V^{*}YV$, with $V$ the magic basis matrix
[see Remark (\ref{magicbasis})]. Then $[Y_{1}, Y_{2}] = 0$ iff 
$B + C$ is normal. Note, while $Y$ is in canonical form, $X$ is not.}
\end{remark} 
 
\section{Conclusions}
In this note, closed form formulae are provided for exponentials
of several important anti-Hermitian $\times 4$ matrices.
These matrices cover many important applications. 
The basic technique is the isomorphism
between real $4\times 4$ matrices and $H\otimes H$.
We believe that this connection is aptly suited to exploit
the properties of $su(4)$ stemming from its direct sum
decomposition into the real skew-symmetric matrices and the
purely imaginary symmetric matrices. While no claim to the
superiority of the representation of the exponential provided
by this work is made, it is our hope that further research will
yield more applications of these formulae.

\end{document}